\documentclass[prl,showpacs,twocolumn,superscriptaddress]{revtex4}
\usepackage{graphicx}

\input{tcilatex}
\begin{document}

\title{Anomalous fluctuations of $s$-wave reduced neutron widths of $%
^{192,194}$Pt resonances}
\author{P.~E.~Koehler}
\thanks{corresponding author}
\affiliation{Physics Division, Oak Ridge National Laboratory, Oak Ridge, TN 37831, USA}
\author{F.~Be\v{c}v\'{a}\v{r}}
\affiliation{Charles University, Faculty of Mathematics and Physics, 180 00 Prague 8,
Czech Republic}
\author{M. Krti\v{c}ka}
\affiliation{Charles University, Faculty of Mathematics and Physics, 180 00 Prague 8,
Czech Republic}
\author{J.~A.~Harvey}
\affiliation{Physics Division, Oak Ridge National Laboratory, Oak Ridge, TN 37831, USA}
\author{K.~H.~Guber}
\affiliation{Nuclear Science and Technology Division, Oak Ridge National Laboratory, Oak
Ridge, TN 37831, USA}
\date{\today }

\begin{abstract}
We obtained an unprecedentedly large number of \textit{s}-wave neutron
widths through $\mathcal{R}$-matrix analysis of neutron cross-section
measurements on enriched Pt samples. Careful analysis of these data rejects
the validity of the Porter-Thomas distribution with a statistical
significance of at least 99.997\%.
\end{abstract}

\maketitle

Neutron resonance parameters remain some of the most important information
for testing random matrix theory (RMT) \cite{Gu98}, even more than fifty
years after such data served as the original impetus for its creation.
Today, RMT pervades the physics of virtually all complex systems and, in the
nuclear physics arena, is most often invoked in studies of quantum chaos 
\cite{We2009}. Given the conserved symmetries involved, RMT for the Gaussian
orthogonal ensemble (GOE) of matrices is expected to correctly describe
fluctuation properties of nuclear levels at relatively high excitation such
as near the neutron threshold. It implicitly assumes that reduced neutron
widths $\Gamma _{\lambda \mathrm{n}}^{0}$ of $s$-wave resonances $\lambda $
follow a Porter-Thomas distribution (PTD) \cite{Po56}, which had been
anticipated before RMT emerged. Currently, the overwhelming consensus is
that $\Gamma _{\lambda \mathrm{n}}^{0}$ data agree with the PTD. However,
there are problems with both the data and analysis techniques used in
reportedly the best test of the PTD to date \cite{Bo83} that call these
results into question \cite{Ko2009}. Measurement and analysis techniques
have improved considerably since then, so it is worthwhile to perform new
tests of the PTD.

In this Letter, we show that rich $\Gamma _{\lambda \mathrm{n}}^{0}$ data
extracted from high resolution neutron total and capture cross sections of $%
^{192,194}$Pt measured at the Oak Ridge Electron Linear Accelerator (ORELA)
facility display a~significant departure from the PTD. To our knowledge,
this result represents the most stringent test of the PTD to date, and the
observed disagreement could have far-reaching consequences.

Loosely stated, the PTD is based on the assumptions that $s$-wave neutron
scattering is a single-channel process, the widths are statistical, and
time-reversal invariance holds; hence, an observed departure from the PTD
implies that one or more of these assumptions is violated, and so could be
very interesting.

To make reliable conclusions regarding the validity of the PTD, it is
important that the data set be as pure, complete, and large as possible.
Perennial problems with neutron resonance data have been (i) contamination
of \textit{s}- by \textit{p}-wave resonances and/or resonances of
neighboring isotopes, (ii) obtaining enough resonances with known spin $J$
to perform statistically meaningful tests, and (iii) missed resonances due
to finite experimental threshold.

In the present case, problem (i) was minimized because Pt is near the peak
of the \textit{s}- and minimum of the \textit{p}-wave neutron strength
functions ($S_{0}/S_{1}\approx 10$), and because we made high resolution
cross-section measurements on natural Pt and four samples enriched in $%
^{192} $Pt, $^{194}$Pt, $^{195}$Pt and $^{196}$Pt.

Problem (ii) was addressed by combining data for two target nuclei $^{192}$%
Pt and $^{194}$Pt, containing the largest number of resonances (158 and 411,
respectively), and by the fact that all $s$-wave resonances have spin $J=1/2$
for these even-even nuclides.

Finally, as described below, the novel approach of using an energy-dependent
threshold in the analysis helps to solve all three problems. As a result,
our $^{192,194}$Pt data are at least as pure, complete, and large as all
previous $\Gamma _{\lambda \mathrm{n}}^{0}$ data which have been used for
testing the PTD.

Details of the measurements can be found in Ref. \cite{Ko2002}. The ORELA
was operated at a pulse rate of 525 Hz, a~pulse width of 8 ns, and a power
of 7-8 kW. Capture measurements were made at a~source-to-sample distance of
40.12 m with a~pair of C$_{6}$D$_{6}$ detectors using the
pulse-height-weighting technique, and were normalized via the saturated
4.9-eV resonance in the $^{197}$Au(n,$\gamma $) reaction. Total neutron
cross sections were measured on a separate flight path via transmission
using a $^{6}$Li-loaded glass scintillator at a source-to-detector distance
of 79.83 m.

The $\mathcal{R}$-matrix code \textsc{sammy} \cite{La2008} was used to fit
both our transmission and capture data and extract resonance parameters.
Resonance energies $E_{\lambda }$ and neutron widths $g_{J}\Gamma _{\lambda 
\mathrm{n}}$, where $g_{J}$ is the statistical factor for resonances with
spin $J$ and $\Gamma _{\lambda \mathrm{n}}$, were used in the subsequent
analysis described below. For even-even targets, $s$-wave resonances have $%
g_{J}\equiv g_{1/2}=1$, and hence, $g_{J}\Gamma _{\lambda \mathrm{n}}=\Gamma
_{\lambda \mathrm{n}}$.

\begin{figure}[b]
\includegraphics[clip,width=1.0\columnwidth]{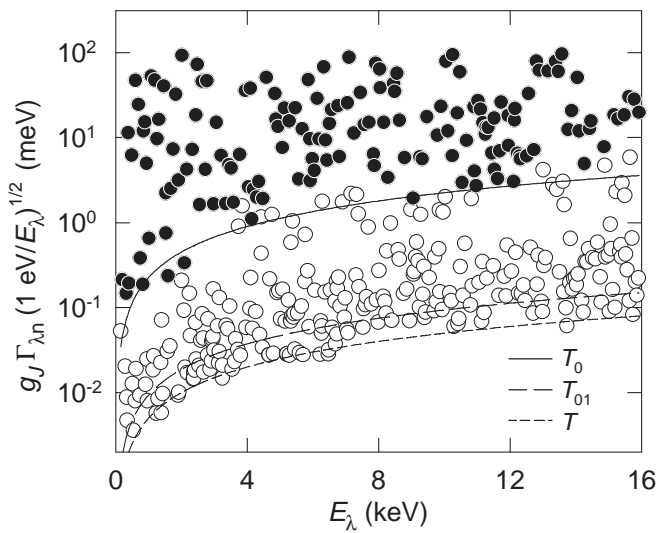} 
\vspace*{-0.3cm}
\caption{Energy-reduced values of observables $g_{J}\,\Gamma _{\protect%
\lambda \mathrm{n}}$ for individual resonances of $^{194}$Pt. Values
represented by full-circle points belong to resonances to which we assigned $%
\ell _{\mathrm{n}}=0$. For the meaning of thresholds $T_{0}$, $T_{01}$ and $%
T $ see the text.}
\label{RedWidVsE}
\end{figure}

An asymmetrical shape in the transmission data could be used to assign $\ell
_{\mathrm{n}}=0$ resonances \cite{Ko2002}, see Fig.~\ref{RedWidVsE}.
However, there remained many weak resonances, most of which are $p$ wave,
for which we could not unambiguously determine the $\ell _{\mathrm{n}}$
value. As shown below, the potential problem posed by these resonances has
been surmounted.%
%

The need to use a threshold on observables $g_{J}\Gamma _{\lambda \mathrm{n}%
} $, to guard against possible systematic errors due to instrumental effects
and \textit{p}-wave contamination, was realized very early on \cite{Po56} in
using such data to test theory. Use of an $E_{\mathrm{n}}$-dependent $s$%
-wave threshold of the form $T_{0}=a\,_{0}E_{\mathrm{n}}^{3/2}$, where $%
a_{0} $ is a~constant factor, is a key improvement in our method compared to
previous analyses (in which only energy-independent thresholds were used)
for at least three reasons. First, \textit{p}-wave contamination is
eliminated equally effectively at all energies. Second, as shown in Fig. \ref%
{RedWidVsE}, our instrumental threshold $T$ follows very closely this same
energy dependence; thus, possible diffusiveness of the instrumental
threshold can be surmounted equally effectively at all energies. Third,
statistical precision of the analysis is maximized by allowing the largest
number of \textit{s}-wave resonances to be included.

The PTD is a special case (degrees-of-freedom $\nu =1$) of the family of $%
\chi ^{2}$ distributions, the probability density function (PDF) of which is
denoted hereafter as $f(x|\nu )$. Because theory predicts fluctuations, but
not average values, it is necessary to scale the data to their average, or
expectation, value; $x\rightarrow \Gamma _{\lambda n}^{0}/\mathrm{E}[\Gamma
_{\lambda n}^{0}]$, where $\mathrm{E}[\bullet ]$ denotes the expectation
value operator.

We began our fluctuation analysis by employing the maximum-likelihood (ML)
method, which has been used since the earliest tests of the PTD, to estimate 
$\nu $ and $\mathrm{E}[\Gamma _{\lambda \mathrm{n}}^{0}]$. We used threshold 
$T_{0}$ shown in Fig. \ref{RedWidVsE} (with factor $a_{0}$ given in Table %
\ref{MLTable}), which was chosen to reduce \textit{p}-wave contamination to
very low levels.

\begin{figure}[b]
\includegraphics[clip,width=0.85\columnwidth]{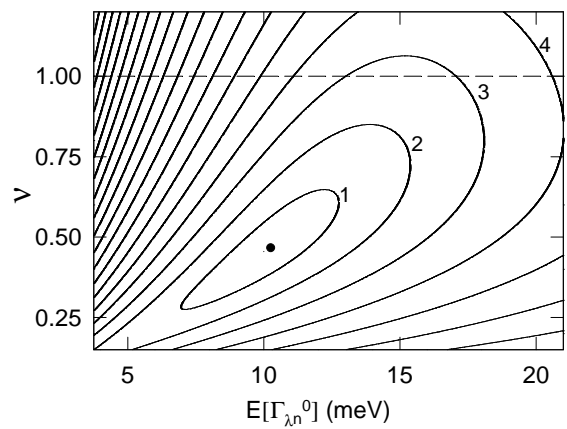} \vspace*{%
-0.3cm}
\caption{Plot of $z\!\left( \protect\nu ,\mathrm{E}[\Gamma _{\protect\lambda 
\mathrm{n}}^{0}]\right) $ constructed from the $^{194}$Pt data.}
\label{ML2DPlot}
\end{figure}

The joint PDF for statistical variables $\Gamma _{\lambda \mathrm{n}}^{0}$
and $E_{\lambda }$ is defined in a~2D region ${\mathcal{I}}$ given by
inequalities $E_{\lambda }<E_{\mathrm{max}}$ and $\Gamma _{\lambda \mathrm{n}%
}^{0}>T_{0}(E_{\lambda })$, where $E_{\mathrm{max}}$ is an~upper limit of
energies $E_{\lambda }$. The expression for this PDF reads 
\begin{equation}
h^{0}\!\left( E_{\lambda },\Gamma _{\lambda \mathrm{n}}^{0}\,|\,\nu ,\mathrm{%
E}[\Gamma _{\lambda \mathrm{n}}^{0}]\right) =Cf\!\left( \left. \frac{\Gamma
_{\lambda \mathrm{n}}^{0}}{\mathrm{E}[\Gamma _{\lambda \mathrm{n}}^{0}]}%
\right\vert \nu \right) .
\end{equation}%
%
%
%
%
%
%
%
%
%
%
%
%
%
The factor $C$, ensuring a~unit norm of $h^{0}$, is $\nu $- and $\mathrm{E}%
[\Gamma _{\lambda \mathrm{n}}^{0}]$-dependent. The ML function was
calculated from all $n_{0}$ pairs $\left[ E_{\lambda _{i}}^{\mathrm{\,exp}%
},\Gamma _{\lambda _{i}\mathrm{n}}^{\mathrm{\,exp}}\right] $, obtained from
the experiment, which fall into the region $\mathcal{I}$. Specifically, 
\begin{equation}
L\left( \nu ,\mathrm{E}[\Gamma _{\lambda \mathrm{n}}^{0}]\right)
=\prod_{i=1}^{n_{0}}h^{0}\!\left( E_{\lambda _{i}}^{\mathrm{\,exp}},\Gamma
_{\lambda _{i}\mathrm{n}}^{\mathrm{0\,exp}}\,|\,\nu ,\mathrm{E}[\Gamma
_{\lambda \mathrm{n}}^{0}]\right) .
\end{equation}%
%
%
%
%
%
%
%
%
%
%
%
%
%
A~contour plot of this ML function in the form%
\begin{equation}
z\!\left( \nu ,\mathrm{E}[\Gamma _{\lambda \mathrm{n}}^{0}]\right) =2^{\frac{%
1}{2}}\left[ \ln L_{\mathrm{max}}-\ln L\left( \nu ,\mathrm{E}[\Gamma
_{\lambda \mathrm{n}}^{0}]\right) \right] ^{\frac{1}{2}}
\end{equation}%
%
%
%
%
%
%
%
%
%
%
%
%
%
is depicted in Fig.~\ref{ML2DPlot}. Here, $L_{\mathrm{max}}$ is the maximum
of the ML function. 
Results of the ML analysis are listed in Table \ref{MLTable}.

\begin{table*}[tbp] \centering%
\caption{Results of ML-based fluctuation analysis of $s$-wave reduced
neutron widths of $^{192,194,196}$Pt resonances.\label{MLTable}} 
\begin{tabular}{cccccccccc}
\hline\hline
Sample & $E_{\max }$ & $a_{0}$ & $n_{0}$ & $\widehat{\nu }_{\mathrm{exp}}$ & 
$a_{01}$ & $\widehat{\mu _{0}}$ & $\widehat{\mu _{1}}$ & $R$ & $S$ \\ 
& (keV) & (eV$^{-\frac{1}{2}}$) &  &  & (eV$^{-\frac{1}{2}}$) & (meV) & (meV)
&  &  \\ \hline
$^{192}$Pt & 4.98 & 7.00$\times $10$^{-8}$ & 153 & 0.57$_{-0.15}^{+0.16}$ & -
& 5.91$\pm $0.71 & (2.68$\pm $0.47)$\times $10$^{-6}$\textbf{%
\footnotemark[1]%
} & 26.1$\pm $4.5 & 0.9970 \\ 
$^{194}$Pt & 15.93 & 2.25$\times $10$^{-7}$ & 161 & 0.47$_{-0.18}^{+0.19}$ & 
9.4$\times $10$^{-9}$ & 14.9$\pm $1.5 & (8.84$\pm $1.09)$\times $10$^{-6}$ & 
25.4$\pm $3.1 & 0.9975 \\ 
$^{196}$Pt & 15.99 & 3.19$\times $10$^{-7}$ & 68 & 0.60$_{-0.26}^{+0.28}$ & 
9.4$\times $10$^{-9}$ & 39.7$\pm $6.4 & (14.8$\pm $1.6)$\times $10$^{-6}$ & -
& - \\ \hline\hline
\end{tabular}
\footnotetext[1]{Due to a high instrumental threshold, $\widehat{\mu_1}$  was deduced assuming that $^{192,194,196}$Pt share a common true value of $\mu_0/\mu_1$.}
\end{table*}%

If $\ln L$ displays near its maximum a~shape close to a~paraboloid,
a~contour for a~fixed value $z=k$ will encircle \textit{approximately} the $%
k\sigma $ confidence region of the ML estimates $\widehat{\mathrm{E}[\Gamma
_{\lambda \mathrm{n}}^{0}]}$ and $\widehat{\nu }$ (referred to hereafter as $%
\widehat{\nu }_{\mathrm{exp}}$). In this way, we found that the difference $%
1-\widehat{\nu }_{\mathrm{exp}}$ equals approximately $2.7\,\sigma $ for $%
^{192,194}$Pt and $1.4\,\sigma $ for $^{196}$Pt. The results for $^{192,194}$%
Pt indicate that the PTD is excluded with high confidence. The result for $%
^{196}$Pt is in agreement with $^{192,194}$Pt, but with reduced statistical
significance owing to the smaller number of resonances observed for this
nuclide. For this reason, we did not include $^{196}$Pt in the subsequent
fluctuation analysis described below.

In the spirit of classical statistics, parameters $\nu $ and $\mathrm{E}%
[\Gamma _{\lambda \mathrm{n}}]$ are not random variables. Consequently, the
function $z$ refers in ideal conditions to a~distribution of \textit{%
estimates} of these parameters, not to the parameters themselves. Further,
as can be deduced from Fig.~\ref{ML2DPlot}, the shape of function $\ln
L\left( \nu ,\mathrm{E}[\Gamma _{\lambda \mathrm{n}}^{0}]\right) $ differs
strongly from a paraboloid, which is also the case for $^{192}$Pt. So, at
this point it is premature to draw a~reliable conclusion from the values of $%
\widehat{\nu }_{\mathrm{exp}}$ for $^{192,194}$Pt.

\begin{figure}[t]
\includegraphics[clip,width=1.00\columnwidth]{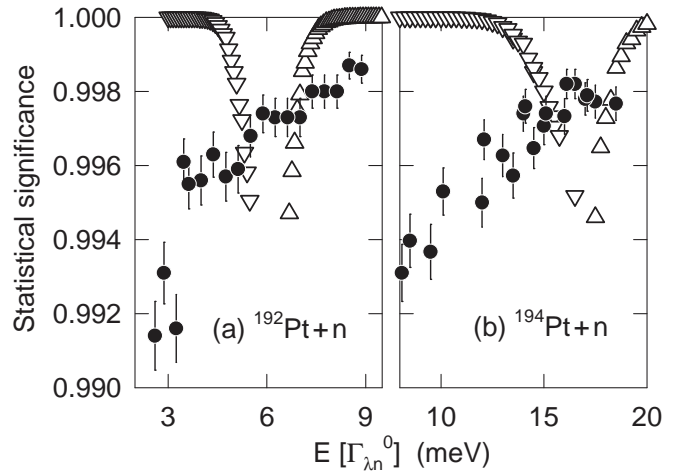} 
\vspace*{-0.3cm}
\caption{Values of probabilities $P(\protect\widehat{\protect\nu }>\protect%
\widehat{\protect\nu }_{\mathrm{exp}})$, $P\left( Z\!>\!Z_{\mathrm{exp}%
}\right) $ and $P\left( Z^{2}\!<\!Z_{\mathrm{exp}}^{2}\right) $ as functions
of $\mathrm{E}[\Gamma _{\protect\lambda \mathrm{n}}^{0}]$ deduced from data
on neutron resonances of $^{192,194}$Pt. Values referring to statistics $%
\protect\nu $, $Z^{2}$ and $Z$ are plotted by symbols {\protect\large $%
\bullet $}, $\bigtriangledown $ and $\triangle $, respectively.}
\label{StatSigVsGn0Ave}
\end{figure}

To check the veracity of the ML results, we undertook additional analyses.
First, for a~given target and a~fixed value $\mathrm{E}[\Gamma _{\lambda 
\mathrm{n}}^{0}]$ we drew from the distribution governed by the PDF, $%
h^{0}\!\left( E_{\lambda },\Gamma _{\lambda \mathrm{n}}^{0}\,|\,\nu \!=\!1,%
\mathrm{E}[\Gamma _{\lambda \mathrm{n}}^{0}]\right) $, a~random sample,
consisting of $n_{0}$ pairs $[E_{\lambda },\Gamma _{\lambda \mathrm{n}}^{0}]$%
. Then, with the aid of the ML analysis, we obtained an~estimate $\widehat{%
\nu }$ for this sample. From a~large number of such samples we constructed
the empirical cumulative distribution function (CDF) of estimates $\widehat{%
\nu }$ and determined a~probability $p_{\nu }=P(\widehat{\nu }>\widehat{\nu }%
_{\mathrm{exp}}\,|\,\mathrm{E}[\Gamma _{\lambda \mathrm{n}}^{0}])$ that 
\textit{a~simulated} value $\widehat{\nu }$ is higher than the corresponding 
\textit{experimental} value $\widehat{\nu }_{\mathrm{exp}}$ listed in Table~%
\ref{MLTable}. Given the value of $\mathrm{E}[\Gamma _{\lambda \mathrm{n}%
}^{0}]$, the probability $p_{\nu }$ represents the statistical significance
at which the validity of the PTD can be rejected. Values $p_{\nu }$ obtained
are plotted in Fig.~\ref{StatSigVsGn0Ave} for $^{192,194}$Pt. They are very
high, but vary considerably with $\mathrm{E}[\Gamma _{\lambda \mathrm{n}%
}^{0}]$. Therefore, we undertook further analyses to impose limits on $%
\mathrm{E}[\Gamma _{\lambda \mathrm{n}}^{0}]$.

To understand how this was achieved, consider a set of $n_{0}$ independent
variables $\{g_{i}\}$ distributed normally with zero mean and unit variance.
For such a set, the statistics 
\begin{equation}
Z=n_{0}^{-\frac{1}{2}}\sum\limits_{i=1}^{n_{0}}g_{i}\;\;\mathrm{and}%
\;\;Z^{2}=\sum\limits_{i=1}^{n_{0}}(g_{i})^{2},
\end{equation}%
%
%
%
%
%
%
%
%
%
%
%
%
%
will be governed by the same normal distribution, and by a $\chi ^{2}$
distribution with $\nu =n_{0}$ degrees of freedom, respectively. Therefore,
for a set of values $\{g_{i}^{\exp }\}$ deduced in an appropriate manner
from experiment, it is straightforward to use $Z$ and $Z^{2}$ to calculate
probabilities for rejecting the null hypothesis that this set is consistent
with the mentioned normal distribution.

To employ these statistics, variable $\Gamma _{\mathrm{n}}^{0}$ \cite{FN1}
needs to be transformed to variable $g$ obeying the considered normal
distribution. This was accomplished in two steps. First, the marginal CDF
for the above-threshold widths 
\begin{equation}
H_{\Gamma _{\mathrm{n}}^{0}}^{0}\!\left( \Gamma _{\mathrm{n}}^{0}|\mathrm{E}%
[\Gamma _{\mathrm{n}}^{0}]\right) =\!\!\!\!\int\limits_{0}^{\Gamma _{\mathrm{%
n}}^{0}}\!\!\!\quad \int\limits_{0}^{E_{\mathrm{max}}}\!\!\!\!h^{0}\!\left(
E_{,}\,\Gamma _{\mathrm{n}}^{0\prime }|1,\mathrm{E}[\Gamma _{\mathrm{n}%
}^{0}]\right) dE\,d\Gamma _{\mathrm{n}}^{0\prime },
\end{equation}%
%
%
%
%
%
%
%
%
%
%
%
was used to transform $\Gamma _{\mathrm{n}}^{0}$ to the variable $r$, which
follows a uniform distribution. This was achieved by substitution $%
r=H_{\Gamma _{\mathrm{n}}^{0}}^{0}(\Gamma _{\mathrm{n}}^{0}\,|\,\mathrm{E}%
[\Gamma _{\mathrm{n}}^{0}])$. In the second step, with the aid of the
inverse CDF of the normal distribution with zero mean and unit variance, $%
G^{-1}(r)$, we made a transformation $r\rightarrow g$, specifically by $%
g=G^{-1}(r)$. Then, quantities $r_{i}=H_{\Gamma _{\mathrm{n}%
}^{0}}^{0}(\Gamma _{\lambda _{i}\mathrm{n}}^{0\,\mathrm{exp}}\,|\,\mathrm{E}%
[\Gamma _{\lambda \mathrm{n}}^{0}])$ for $i=1,2,\dots n_{0},$ and hence sets 
$\{g_{i}^{\exp }\}$ were calculated using values $\Gamma _{\lambda _{i}%
\mathrm{n}}^{0}>T_{0}(E_{\lambda _{i}})$ from the experiments. Following
this procedure, we easily determined probabilities $p_{Z}=P\left( Z\!>\!Z_{%
\mathrm{exp}}\,|\,\mathrm{E}[\Gamma _{\lambda \mathrm{n}}^{0}]\right) $ and $%
p_{Z^{2}}=P\left( Z^{2}\!<\!Z_{\mathrm{exp}}^{2}\,|\,\mathrm{E}[\Gamma
_{\lambda \mathrm{n}}^{0}]\right) $. Here, for a fixed value $\mathrm{E}%
[\Gamma _{\lambda \mathrm{n}}^{0}]$, values of $Z_{\mathrm{exp}}$ and $Z_{%
\mathrm{exp}}^{2}$ are given by Eqs. (4) after replacement $g_{i}\rightarrow
g_{i}^{\exp }$. Hence, probabilities $p_{Z}$ and $p_{Z^{2}}$ represent
separate statistical significances for rejecting the PTD at various values
of $\mathrm{E}[\Gamma _{\lambda \mathrm{n}}^{0}]$. Both probabilities,
calculated from $^{192,194}$Pt data, are plotted in Fig. \ref%
{StatSigVsGn0Ave}. As seen, the critical values of $\mathrm{E}[\Gamma
_{\lambda \mathrm{n}}^{0}]$ for testing the validity of the PTD range from
5.3 to 6.9 meV and from 15.0 to 18.3 meV for $^{192}$Pt and $^{194}$Pt,
respectively.%

\vspace*{0cm} Before proceeding further, we performed an additional analysis
to verify that $p$-wave contamination was negligibly small. To this end, we
performed ML calculations based on a~hybrid PDF for a~mixture of $s$ and $p$
observables $g_{J}\Gamma _{\lambda \mathrm{n}}$ obeying two separate PTDs: 
\[
h^{01}\!\left( E_{\lambda },\,g_{J}\Gamma _{\lambda \mathrm{n}}\,|\,\mu
_{0},\mu _{1},\theta \right) \!=\!\frac{D}{4}\frac{\mathrm{1eV}^{\frac{1}{2}}%
}{\mu _{0}E_{\lambda }^{\frac{1}{2}}}\,f\!\left( \frac{g_{J}\Gamma _{\lambda 
\mathrm{n}}}{\mu _{0}}\left. \frac{\mathrm{1eV}^{\frac{1}{2}}}{E_{\lambda }^{%
\frac{1}{2}}}\,\right\vert \,1\right) 
\]%
\vspace*{-0.2cm} 
\[
+\theta \,\frac{3D}{4}\frac{\mathrm{1eV}^{\frac{3}{2}}}{\mu _{1}E_{\lambda
}^{\frac{3}{2}}}\,f\!\left( \frac{g_{J}\Gamma _{\lambda \mathrm{n}}}{\mu _{1}%
}\left. \frac{\mathrm{1eV}^{\frac{3}{2}}}{E_{\lambda }^{\frac{3}{2}}}%
\,\right\vert \,1\right) .
\]%
%
%
%
%
%
%
%
%
%
%
Here, $\mu _{0}$ and $\mu _{1}$ stand for expectation values $\mathrm{E}%
[\Gamma _{\lambda \mathrm{n}}^{0}]$ and $\mathrm{E}\left[ g_{J}\,\Gamma
_{\lambda \mathrm{n}}(\mathrm{1eV}/E_{\lambda })^{3/2}\right] $ referring to 
$s$- and $p$-wave resonances, respectively, while $\theta $ represents the $J
$-independent resonance-density ratio $\rho _{J^{-}}/\rho _{J^{+}}$. The PDF 
$h^{01}$ is defined in the region limited from below by~threshold $%
T_{01}=a_{01}E_{\mathrm{n}}^{3/2}$ with $a_{01}\ll a_{0}$ which is at the
same time safely higher than experimental threshold $T=aE_{\mathrm{n}}^{3/2}$%
. We assumed that $\mu _{1}$ is $J$-independent. For mass numbers $A\approx
190$ this is justified, as it holds with about 1\% precision that $%
g_{1/2}/g_{3/2}=f(3/2)/f(1/2)$, where $f(J)$ is the resonance-density spin
factor \cite{Zhon}. Factor $D$ ensures the unit norm of $h^{01}$.

Following a path analogous to that described above, we constructed ML
function $L\left( \mu _{0},\mu _{1}\right) $ and arrived at estimates $%
\widehat{\mu _{0}}$ and $\widehat{\mu _{1}}$ given in Table \ref{MLTable}.
In case of $^{192}$Pt, $\widehat{\mu _{1}}$ has been determined indirectly,
see Table \ref{MLTable}. The method of its determination is supported by the
fact that ratios $\widehat{\mu _{0}}/\widehat{\mu _{1}}$ for $^{194,196}$Pt
are within their rms uncertainties equal each other. In Table \ref{MLTable},
values of the dimensionless, energy-independent quantity $R=(1\mathrm{\ eV}%
^{3/2})\,a_{0}/\widehat{\mu _{1}}$, which represent ratios of $T_{0}(E_{%
\mathrm{n}})$ to~\textit{local} expectation values $\mathrm{E}[\Gamma
_{\lambda \mathrm{n}}]$ for $p$-wave resonances, are listed for $^{192,194}$%
Pt targets. From these values and their rms uncertainties, we
calculated~probabilities of 0.069\% and 0.0047\% for $^{192}$Pt and $^{194}$%
Pt, respectively, that among the observables $g_{J}\Gamma _{\lambda \mathrm{n%
}}>T_{0}(E_{\lambda })$ there occurs one which belongs to a $p$-wave
resonance, thus verifying that $p$-wave contamination is negligible.

With the question of $p$-wave contamination settled then, the data in Fig.~%
\ref{StatSigVsGn0Ave} indicate that for $s$-wave resonances in $^{192}$Pt
and $^{194}$Pt, the validity of the PTD is rejected with statistical
significance levels $S_{192}=0.9970$ and $S_{194}=0.9975$, respectively, in
excellent agreement with our initial ML analysis. Because results for the
two isotopes should be independent, the combined probability that the PTD is
valid is less than $1.2\times 10^{-5}$. Although we have shown above that it
is very unlikely, if a $p$-wave intruder occurs among the $s$-wave $^{192}$%
Pt observables, we calculate that if the resonance having the smallest $%
g_{J}\Gamma _{\lambda \mathrm{n}}$ value (relative to threshold $T_{0}$) is
considered to be $p$ wave, this probability will increase in the worst case
to $2.8\times 10^{-5}$.

We conclude that our data reject the validity of the PTD with a~statistical
significance of at least 99.997\%. This inescapable conclusion has been made
thanks to rich experimental data obtained using state-of-the-art neutron
spectroscopy, and the implementation of a~novel approach for testing the
PTD. On the other hand, equally convincing evidence that the PTD holds for
some or the majority of heavy and intermediate-weight nuclei is still
missing.

This result implies that at least one of the three assumptions behind the
PTD is violated. For energies of the measurements reported herein, only
elastic scattering is possible, so the single-channel assumption is valid.
Also, addition of another channel would result in $\nu >1$, which would
disagree even more strongly with the data than the PTD does. Violation of
time-reversal invariance also implies $\nu >1$, and therefore also is
excluded. Hence, our results indicate the assumption that the widths are
statistical is violated. However, there are no indications of
non-statistical effects such as doorway states in the data.

One possible explanation is suggested by the calculations of Ref. \cite{Al92}
in which it was found that transition strength distributions deviated
further from the PTD, in the direction of smaller $\nu $, as
model~quantum-mechanical systems became more collective. Hence, our result
that $\nu \approx 0.5$ for the $^{192,194}$Pt+n systems suggests the
surprising conclusion that $^{193,195}$Pt display regular, rather than the
expected chaotic, behavior at relatively high excitations near the neutron
threshold.

Alternatively, our results also could be interpreted as indicating that the
PDF for reduced neutron width amplitudes for the $^{192,194}$Pt+n systems
are not form invariant. In Ref. \cite{Kr63} it was shown that this
form-invariance assumption could replace the original \cite{Po56} somewhat
qualitative "statistical" assumption as part of a more general derivation of
the PTD. Violation of this assumption could have far-reaching consequences.


This work was supported by the U.S. Department of Energy under Contract No.
DEAC05-00OR22725 with UT-Battelle, LLC, and by Czech Research Plans
MSM-021620859 and INGO-LA08015.


\begin{thebibliography}{99}
\bibitem{Gu98} T. Guhr \textit{et al}., Phys. Repts. \textbf{299}, 189
(1998).

\bibitem{We2009} H. A. Weidenmuller and G. E. Mitchell, Rev. Mod. Phys. 81,
539 (2009).

\bibitem{Po56} C.~E.~Porter and G.~E.~Thomas, Phys. Rev. \textbf{104}, 483
(1956).

\bibitem{Bo83} O. Bohigas, R. U. Haq, and A. Pandey, in \textit{Nuclear Data
for Science and Technology}, edited by K. H. Bockhoff (D. Reidel, Dodrecht,
1983), p. 809.

\bibitem{Ko2009} P. E. Koehler, EPJ Web of Conferences \textbf{2}, 05001
(2010).

\bibitem{Ko2002} P. E. Koehler \textit{et al}., J. Nucl. Sci. and Tech.,
Suppl. 2, 546 (2002).

\bibitem{La2008} N. M. Larson, Oak Ridge National Laboratory Technical
Report No. ORNL/TM-9179/R8, 2008.

\bibitem{FN1} For the sake of brevity subscripts $\lambda $ are dropped.

\bibitem{Zhon} H.~Zhongfu \textit{et al}., Chin. J. Nucl. Phys. \textbf{13},
147 (1991).

\bibitem{Al92} Y. Alhassid and A. Novoselsky, Phys. Rev. C \textbf{45}, 1677
(1992).

\bibitem{Kr63} T.J. Krieger and C. E. Porter, J. Math. Phys. \textbf{4},
1272 (1963).
\end{thebibliography}
\end{document}